\documentclass{jnmp}

\usepackage{amsmath}

\JNMPnumberwithin{equation}{section}

\theoremstyle{definition}

\begin{document}

%
\renewcommand{\evenhead}{A Constantin and J Lenells}
\renewcommand{\oddhead}{The Inverse Scattering Approach to
  the Camassa-Holm Equation}

%
\thispagestyle{empty}

\FirstPageHead{10}{3}{2003}{\pageref{firstpage}--\pageref{lastpage}}{Letter}

\copyrightnote{2003}{A Constantin and J Lenells}

\Name{On the Inverse Scattering Approach to
  the Camassa-Holm Equation}

\label{firstpage}

\Author{Adrian CONSTANTIN~$^\dag$ and Jonatan LENELLS}

\Address{Department of Mathematics, Lund University, P.O. Box 118, SE-221 00 Lund, Sweden \\
$^\dag$  Corresponding author. ~~E-mail: adrian.constantin@math.lu.se}
\Date{Received November 25, 2002; Revised January 06, 2003;
Accepted January 07, 2003}

\begin{abstract}
\noindent A simple algoritm for the inverse scattering approach to the
Camassa-Holm equation is presented.
\end{abstract}

%
\section{Introduction}

The nonlinear partial differential equation 
\begin{equation}
  m_t+2 u_x+um_x+2mu_x=0,\quad t>0, \  x\in
  \mathbb R,
\end{equation}
in dimensionless space-time variables $(x,t)$ models the
unidirectional propagation of two-dimensional waves in shallow water
over a flat bottom. In (1.1), $u(t,x)$ represents the horizontal
component of the fluid
velocity, or, equivalently, the water's free surface, 
and $m=u-u_{xx}$ is the momentum variable cf. \cite{CH} 
(see also \cite{J1} for an alternative derivation). The solitary 
waves of (1.1) are solitons - they regain
their shape and speed after interacting nonlinearly with other
solitary waves (see \cite{CH} \cite{CS} \cite{J2} \cite{L}). While
some initial profiles evolve into waves of permanent form, others
yield waves that break in finite time \cite{C2} \cite{CE}. Another
aspect of interest of (1.1) is its bi-Hamiltonian structure \cite{FF} and the
induced existence of infinitely many conservation laws. This feature is
connected to the (formal) integrability of the equation,
established in \cite{CH} by finding an isospectral problem associated
to (1.1). 

Physically relevant cases are solutions of (1.1) with a periodic
dependence upon the spatial $x$-variable, as well as solutions which 
decay at infinity. The objective of this note is to prove the 
integrability of (1.1) for solutions which decay at infinity (the
periodic case is treated in \cite{C1} \cite{CM} \cite{CE2}). For 
a discussion of
the scattering problem for (1.1) we refer to \cite{C2} and
\cite{L}. Recently, in \cite{CL}, we proposed an algorithm to solve the
inverse scattering problem for (1.1). Our aim is to present here a
considerable simplification of the approach in \cite{CL} and to show
the applicability of the new approach by means of an example.

\section{The inverse scattering approach}

If the initial momentum $m_0=m(0,\cdot)$ belongs to the Schwartz class
of smooth functions $f:\mathbb R \to \mathbb R$ such that for 
all $n_1,n_2 \ge 0$, $\displaystyle\sup_{x \in \mathbb R} 
|x^{n_1}\partial_x^{n_2}f(x)|<\infty$, and $m_0 +1 > 0$, then both
properties are preserved by the flow of (1.1) cf. \cite{C2}
\cite{CL}. The isospectral 
problem in $L^2(\mathbb R)$ for (1.1) is (see \cite{CH})
\begin{equation}
  \psi_{xx}=\frac{1}{4}\,\psi + \lambda (m+ 1) \psi
\end{equation}
with continuous spectrum $(-\infty, -
\frac{1}{4}]$
and at most finitely many eigenvalues in the interval
 $(-\frac{1}{4},0)$ cf. \cite{C2}. The Liouville transformation
$$
\varphi(y)= \Bigl(m(x)+1\Bigr)^\frac{1}{4}\, \psi(x)\quad\hbox{where}
\quad y=x+\int_{-\infty}^x \Bigl[\sqrt{m(\xi)+ 1}\,-1\Bigr]\, d\xi
$$ 
converts (2.1) into
\begin{equation}
-{{d^2 \varphi} \over {d y^2}}+ Q \varphi=\mu \varphi.
\end{equation}
Here
\begin{equation}
Q(y)={1 \over {4q(y)}}+{{q_{yy}(y)} \over{4\,q(y)}}-
{{3\, q_y^2(y)} \over {16\, q^2(y)}} \,
 - \, {1 \over {4}}
\end{equation}
with $q(y)=m(x)+1$ and spectral parameter $\mu=-{1 \over
  {4}}-\lambda$. Suitable scattering data for (2.1) happen to be the
usual scattering data for the problem (2.2), which evolve linearly at
constant speed under the Camassa-Holm flow cf. \cite{CL}. Therefore 
the classical Marchenko approach (see \cite{DJ}) is applicable and finding
$Q(t,y)$ amounts to solving a linear integral equation determined by
the scattering data for $Q_0(y)$, data available from the
knowledge of $m_0(x)$. The only intricate point of this approach is
the recovery of $m(t,x)$ from $Q(t,y)$. This requires us to solve, given
$Q$, the nonlinear second order differential equation (2.3) for $q$, 
and then to perform the
coordinate transform $y \mapsto x$. Equation (2.3) is a Pinney
equation \cite{P} but the solution for $q$, given
$Q$, obtained in \cite{P} is not convenient for our purposes (this
approach was used in \cite{C2} and leads to unnecessary
complications). A quite intricate (but
nevertheless effective) algorithm
for the recovery of $m$ was proposed in \cite{CL}. Below we present 
an alternative
  approach that gives a more direct and less complicated
  solution. For convenience we drop the time-dependence in 
the formulation.\bigskip

{\bf Theorem} {\it Let $f(y)$ be the Jost function at $y=\infty$ for
  the eigenvalue equation
\begin{equation}
\varphi_{yy}=(Q + \frac{1}{4})\,\varphi,
\end{equation} 
i.e. $f$ is the unique solution of $(2.4)$ with the asymptotic behavior
$$f(y) \approx e^{-y/2}\quad and \quad 
 f'(y) \approx -\,\frac{1}{2}\,e^{-y/2}\quad as \quad y \to \infty.$$
If $H:\,\mathbb R \to \mathbb R$ is the bijection given by
$H(y)=\displaystyle\int_{-\infty}^y\frac{d\xi}{f^2(\xi)}$, then}
\begin{equation}
m(x)+1=e^{2x}\,f^4(H^{-1}(e^x)),\quad x \in \mathbb R.
\end{equation}

\smallskip
\noindent
{\it Proof.} Observe that (2.4) is precisely (2.2) with
$\mu=-\,\frac{1}{4}$. Hence $\lambda=0$, and the Liouville
transformation maps (2.2) into 
$$\psi_{xx}=\frac{1}{4}\,\psi,$$
where $\varphi(y)= \Bigl(m(x)+1\Bigr)^\frac{1}{4}\psi(x)$. The above
equation has the solution $e^{-x/2}$ so that (2.2) has the
corresponding solution
$h(y)=\Bigl(m(x)+1\Bigr)^\frac{1}{4}e^{-x/2}$. But one can easily
check that the function $h$ has
the asymptotic behavior at $y=\infty$ prescribed for $f$, so $f=h$, i.e. 
$f(y)=q^{1/4}(y)\,e^{-x/2}$. This means that
$\displaystyle\frac{dy}{f^2(y)}=\frac{e^x\,dy}{\sqrt{q(y)}}=e^x\,dx$.
Therefore $H(y)=\displaystyle\int_{-\infty}^x e^s\,ds=e^x$ is clearly
invertible and $y=H^{-1}(e^x)$. From the relation
$m(x)+1=q(y)=f^4(y)e^{2x}$ we now obtain (2.5).$\quad\Box$\bigskip

{\bf Remark} The previous result reduces the recovery of $m(x)$ from
$Q(y)$ to solving the linear integral equation
$$f(y)=e^{-y/2}+\int_y^\infty
\Bigl(e^{(\xi-y)/2}-e^{(y-\xi)/2}\Bigr)\,Q(\xi)\,f(\xi)\,d\xi,\qquad y
\in \mathbb R,$$
and computing the inverse of the function
$H(y)=\displaystyle\int_{-\infty}^y\frac{d\xi}{f^2(\xi)}$.$\quad\Box$\bigskip

{\bf Example} A solitary wave for (1.1) is a 
solution $m(t,x)=\Phi(x-ct)$ with a
profile $\Phi$ that decays at infinity. Solitary waves can exist only
for speeds $c>2$, but no mathematical expression in closed form seems
to be available for $\Phi$ (see \cite{CS}), despite the fact 
that the corresponding
potential $Q(y)$ is given explicitly by  
$$Q(y)=\displaystyle{-\,\frac{c-2}{2c\,\cosh^2\Bigl(
\sqrt{\frac{c-2}{4c}}\,(y-y_0)\Bigr)}},\qquad y \in \mathbb R,$$
with $y_0 \in \mathbb R$ 
(see \cite{L}). If we choose for simplicity $y_0=0$ and $c=8/3$, then
$$Q(y)+\frac{1}{4}=\frac{2 \cosh^2(y/4)-1}{8\cosh^2(y/4)},\qquad y 
\in \mathbb R.$$
Note that $g(y)=4\,\cosh^2(y/4)$ is a solution to (2.4). But then $y
\mapsto g(y)\displaystyle\int_y^\infty\frac{d\xi}{g^2(\xi)}$ is also a
solution to (2.4). Since
$\displaystyle\int_y^\infty\frac{d\xi}{g^2(\xi)}=\frac{6e^{-y/4}+2e^{-3y/4}}{3\cosh^3(y/4)}$,
we deduce that 
$$f(y)=\frac{3 e^{-y/4}+e^{-3y/4}}{6\cosh(y/4)},\qquad y \in \mathbb R,$$
and an exact formula for the profile of the solitary wave
emerges.$\quad\Box$\bigskip

{\bf Acknowledgement} The authors thank the referee for valuable remarks.

\label{lastpage}

\end{document}